# Design and Analysis of RS Sort

**Harsh Ranjan[1], Sumit Agarwal[1] and Niraj Kumar Singh[1]**

[1]**Department of Computer Science & Engineering, B.I.T. Mesra, Ranchi-835215, India Email address of the corresponding author: niraj_2027@yahoo.co.in**

*Abstract:*
This paper introduces a new comparison base stable sorting algorithm, named RS sort. RS Sort involves only the comparison of pair of elements in an array which ultimately sorts the array and does not involve the comparison of each element with every other element. RS sort tries to build upon the relationship established between the elements in each pass. Suppose there is an array containing three elements $a_1$, $a_2$, $a_3$ and if a relationship exist such that $a_1<a_2$ and $a_2<a_3$ then it can be established that $a_1<a_3$ and so there is no need to compare $a_1$ and $a_3$. Sorting is a fundamental operation in computer science. RS sort is analyzed both theoretically and empirically. We have performed its Empirical analysis and compared its performance with the well-known quick sort for various input types.

**Keywords: -** algorithm, quick sort, RS sort, theoretical analysis, empirical analysis, uniform distribution model, poisson model, binomial model,

1. INTRODUCTION

This paper introduces a new comparison base stable sorting algorithm, named RS sort. Though many sorting algorithms have been developed, no single technique is best suited for all applications. In basic comparison sort algorithm we need to check each element with the rest of the array in order to find its appropriate location. In RS sort we only need to compare selective pairs of whose elements are distant apart in a defined manner. This selective comparison among elements saves a fair amount of time and comparisons.

Although the theoretical worst case complexity of RS sort is $Y_{worst}(n) = O(n\sqrt{n})$, the experimental results reveal that with $O_{emp}(nlgn)^{1.333}$ time complexity for typical inputs it can perform optimally.

2. ALGORITHM: RS SORT

RS Sort involves only the comparison of pair of elements in an array which ultimately sorts the array and does not involve the comparison of each element with every other element. RS sort tries to build upon the relationship established between the elements in each pass. For an input ($a_1$, $a_2$, $a_3$) let $a_1<a_2$ and $a_2<a_3$, then it can be easily inference that $a_1<a_3$ and so there is no need to compare $a_1$ and $a_3$. RS sort uses this technique to place each element in their appropriate location by saving significantly large number of comparisons.

RS sort first determines the minimum length such that all elements get placed in their appropriate locations. This length refers to the maximum forward distance a particular index can be compared with. It starts with this length and goes down to one at each point comparing every element only with one element that is a fixed length forward to it. This minimum value of length can be easily found out by binary search as for all values greater than this length the array will be sorted and for all values less than it, it will be partially sorted. If there is an array of four elements then initially $a_1$ needed to be compared with $a_2$, $a_3$, $a_4$ and same goes for $a_2$, $a_3$ and $a_4$ before completing the sorting but if length equals two then in first loop $a_1$ is compared only with $a_3$ and $a_2$ is compared only with $a_4$ and in the second loop when length decrements by one $a_1$ gets compared only with $a_2$, $a_2$ gets compared only with $a_3$ and $a_3$ gets compared only with $a_4$. These comparisons takes place one by one and at each point the value at an index might change and the updated value at that index gets used for future comparisons. Thus in only five comparisons when length equals to two and three when length equals to one we have sorted the entire array instead of a total of twelve. This differences increases greatly as the size of array increases.

This minimum length does not have a general formula which can be given for all input size but a rough estimate can be made which gives the minimum value for most of the cases and for few cases it gives a slightly higher value which ultimately does the sorting job perfectly. Let us denote this minimum values by *K*. Then *K=T \*lgn* where $T = \left\lceil \frac{|\sqrt{1+8n}-1|}{2*lgn} \right\rceil$

*Derivation of T:* Let n denote the input size of a sample and h equals to *lgn*. Maximum jump required by any element to go to its correct position = *n-1* (smallest element is at the last position or largest at first.) After x iteration maximum jump that can be made by any element from its given location by RS sort is *1+2+3+…+x*. Multiplying *x* by *h* and summing the above series we get *(x\*h)\*(x\*h+1)/2*. Now this value needs to be greater than *n-1* so that every element can reach its appropriate location in worst case. On comparing them: *(x\*h)\*(x\*h+1)/2≥n* (Replacing *n* by *n-1* for calculation ease.) Considering *x\*h=z*, we have: *z\*(z+1) ≥ 2\*n*

⇨ $z^2 + z - 2*n \geq 0$
⇨ $z = (-1+\sqrt{1+8n})/2$ , and since *z=x\*h*, we get

$x = \sqrt{[(1+8n)-1]}/2lgn.$

Thus *T=[x]*, for covering boundary cases at some places. On solving this quadratic relation for x since h is a constant gives the required formula for *T* as $T = \left\lceil \frac{|\sqrt{1+8n}-1|}{2*lgn} \right\rceil$. The minimum length is given as *T\*lgn*. It can be seen that in general case any length less than this can't sort the array totally as each element would not end up at their appropriate location and every length greater will. RS sort is analyzed both theoretically and empirically. We have done theoretical analysis to get its worst case performance in terms of big-oh notation. Average case analysis is done using statistical

```
Algorithm: RS_Sort (A, n):
//Here A [0: n-1] is the input array of size n
    FOR (I=T*[lgn]; I≥1; I=I-1) // T = ⌈(√(1+8n)−1)/(2*lgn)⌉
        FOR (J=0; J+I< n; J=J+1)
            IF (A[J]>A [J+I])
                Exchange (A[J], A [J+I])
```

## 2.1 ANALYSIS OF RS SORT
bound estimate (also called empirical-O).

### 2.1.1 Theoretical Analysis (Worst case only)
Referring to the pseudo code (it contains two for loops) of RS sort its runtime complexity is expressed as the following summation equation:

$$Y(n) = \sum_{i=T\lg n}^{i=1} 1 \sum_{j=0}^{j+i<n}(1)$$

The worst case equation is: $Y_{worst}(n) = O(T*lgn*n) = O(n\sqrt{n})$, which is obtained by substituting $T = \left\lceil \frac{|\sqrt{1+8n}-1|}{2*lgn} \right\rceil$.

### 2.1.2 Empirical Analysis (Average case only)
This section includes empirical results obtained for average case analysis of RS and quick sorts. The algorithm was run for data obtained from various uniform and non-uniform discrete distribution data model like *Uniform distribution*, *Poisson distribution* and *Binomial distribution*. The performance of RS sort is also compared with standard version of quick sort algorithm (Hoare, 1962) for the similar input types. The observed mean time (in sec) of 1000 trials was noted in table (1). Average case analysis was done by directly working on program run time to estimate the weight based statistical bound over a finite range by running computer experiments (Fang et al. 2006; Sacks et al. 1989). This estimate is called empirical O (Chakraborty and Sourabh, 2010; Sourabh and Chakraborty, 2007). Here time of an operation is taken as its weight. Weighing permits collective consideration of all operations into a conceptual bound which we call a statistical bound in order to distinguish it from the count based mathematical bounds that are operation specific. The way we design and analyze our computer experiment has certainly a great impact on the credibility of empirical-O. See reference (Chakraborty and Sourabh, 2010) for more insight into the philosophy behind statistical bound and empirical-O. The statistical analysis and the various interpretations are guided by (Mathews 2010).

The samples are generated randomly, using a random number generating function, to characterize discrete uniform, poisson, and binomial distribution models with $k$, $\lambda$, and $(m, p)$ as its respective parameters. Our sample sizes lie in between $1*10^5$ and $20*10^5$.

Table 1: Observed mean times in second(s) for RS and quick sorts

|  | RS SORT | | | QUICK SORT | | |
| --- | --- | --- | --- | --- | --- | --- |
| N | DU $k = 50$ | Poisson $\lambda = 4$ | Binomial (400, 0.5) | DU $k = 50$ | Poisson $\lambda = 4$ | Binomial (400, 0.5) |
| 100000 | 0.203 | 0.187 | 0.187 | 0.218 | 1.422 | 0.125 |
| 200000 | 0.531 | 0.515 | 0.531 | 0.812 | 5.688 | 0.421 |
| 300000 | 0.969 | 0.937 | 1.000 | 1.811 | 9.883 | 0.938 |
| 400000 | 1.500 | 1.508 | 1.515 | 3.213 | 14.313 | 1.641 |
| 500000 | 2.112 | 2.161 | 2.140 | 5.057 | 19.524 | 2.516 |
| 600000 | 2.793 | 2.915 | 2.828 | 7.299 | 25.139 | 3.657 |
| 700000 | 3.500 | 3.619 | 3.484 | 10.067 | 30.213 | 5.234 |

| 800000 | 4.328 | 4.619 | 4.298 | 13.047 | 37.169 | 6.406 |
| 900000 | 5.141 | 5.301 | 5.121 | 16.364 | 42.130 | 8.110 |
| 1000000 | 6.023 | 6.042 | 5.984 | 20.251 | 46.929 | 10.016 |
| 1100000 | 6.953 | 7.027 | 6.938 | 24.564 | 51.142 | 12.095 |
| 1200000 | 7.924 | 7.975 | 7.933 | 29.095 | 56.521 | 14.438 |
| 1300000 | 8.876 | 9.079 | 8.876 | 34.219 | 60.998 | 16.986 |
| 1400000 | 9.878 | 10.079 | 9.891 | 39.496 | 64.321 | 19.579 |
| 1500000 | 10.997 | 11.108 | 10.969 | 45.355 | 69.032 | 22.501 |
| 1600000 | 12.050 | 12.239 | 12.047 | 51.707 | 74.328 | 25.693 |
| 1700000 | 13.290 | 13.548 | 13.291 | 58.335 | 78.431 | 28.485 |
| 1800000 | 14.441 | 14.774 | 14.449 | 65.279 | 84.320 | 32.563 |
| 1900000 | 15.627 | 15.907 | 15.835 | 72.787 | 90.001 | 36.222 |
| 2000000 | 16.871 | 17.243 | 17.070 | 80.403 | 95.113 | 41.012 |

Below we present two comparative plots for RS against the quick sort. The figures 1&2 reveal the superiority of RS sort for discrete uniform and poisson distribution data models for the specified parameter values.

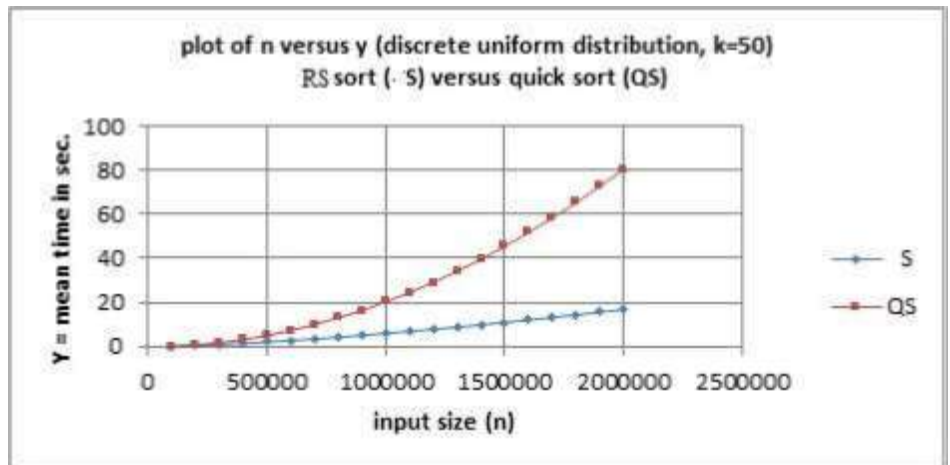

Figure (1): Plot of n versus y (discrete uniform distribution, k=50)

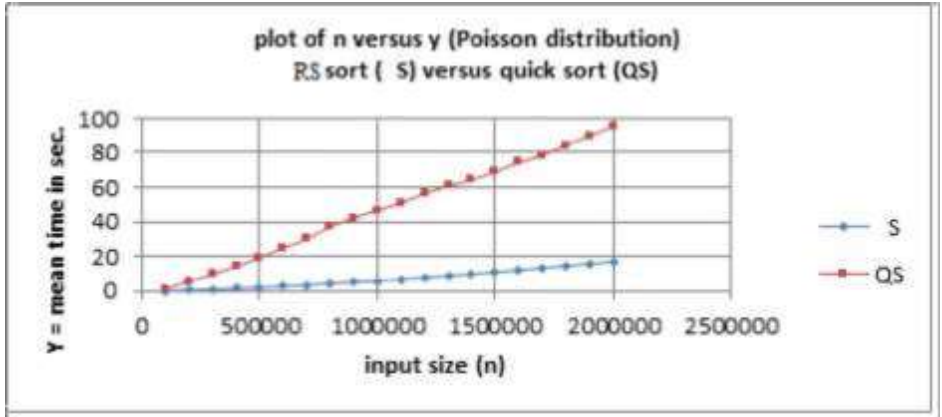

Figure (2): Plot of n versus y (Poisson distribution, $\lambda=4$)

Table 2: General Regression Analysis: Y versus n, nlogn, n^2

```
General Regression Analysis: Y versus n, nlog₂n, n^2:

Box-Cox transformation of the response with specified lambda = 0.75
Regression Equation

Y^0.75  =   0.0383403 - 3.90451e-006 n + 3.88879e-007 nlogn - 4.74249e014
n^2

Coefficients

Term              Coef     SE Coef         T       P
Constant     0.0383403   0.0250803    1.52870   0.146 n
-0.0000039   0.0000010   -3.79282   0.002 nlogn
0.0000004    0.0000001    7.38394   0.000 n^2         -
0.0000000    0.0000000   -1.06788   0.301

Summary of Model

S = 0.0125067       R-Sq = 100.00%       R-Sq(adj) = 100.00%
PRESS = 0.00435678   R-Sq(pred) = 100.00%

Analysis of Variance

Source        DF    Seq SS    Adj SS    Adj MS       F            P
Regression    3    121.313   121.313   40.4378    258523   0.000000    n
1    121.174    0.002    0.0023       14   0.001597    nlogn      1
0.139     0.009    0.0085       55  0.000002    n^2         1     0.000
0.000     0.0002       1   0.301421
Error         16    0.003     0.003    0.0002
Total         19   121.316

Fits and Diagnostics for Unusual Observations for Transformed Response
Obs    Y^0.75      Fit      SE Fit     Residual   St Resid
```

```
Fits for Unusual Observations for Original Response

Obs       Y       Fit
  1    0.203    0.1949    X
 16   12.050   12.1137    R

R denotes an observation with a large standardized residual.
X denotes an observation whose X value gives it large leverage.

Durbin-Watson Statistic:
Durbin-Watson statistic = 1.75208
```

```
  1   0.30243   0.29333   0.0106239    0.0090990    1.37876    X
 16   6.46756   6.49317   0.0043949   -0.0256095   -2.18714    R
```

The program runtime data corresponding to the discrete uniform distribution samples is fitted for a quadratic model of type: $y=b_0+b_1n+b_2nlog_2n+b_3n^2$. The response variable is transformed using Box-Cox transformation (Box and Cox, 1964) technique, with specified lambda equal to 0.75 where it the parameter of this transformation, to get a suitable model. Below is the corresponding regression model (complete result is available in table 2).

Regression Equation:

$$y^{0.75} = 0.0383403 - 3.90451e\text{-}006\ n + 3.88879e\text{-}007\ nlogn - 4.74249e\text{-}014\ n^2$$

As the statistical significance of quadratic term is very weak we ignore it from our model. It reduces the resulting model as: $y^{0.75} = 0.0383403 - 3.90451e\text{-}006\ n + 3.88879e\text{-}007\ nlogn$.
Consequently we have $y^{0.75} = O_{emp}(nlog_2n)$, which implies that $y=O_{emp}(nlog_2n)^{1/0.75} = O_{emp}(nlog_2n)^{1.333}$. The standard error of this model is very low (S=0.0125067) and it explains almost all the variations (as R-Sq(adj) value is equal to 100%). These observations led us to conclude that the average case complexity of RS sort is: $Y_{avg}(n) = O_{emp}(nlog_2n)^{1.333}$.

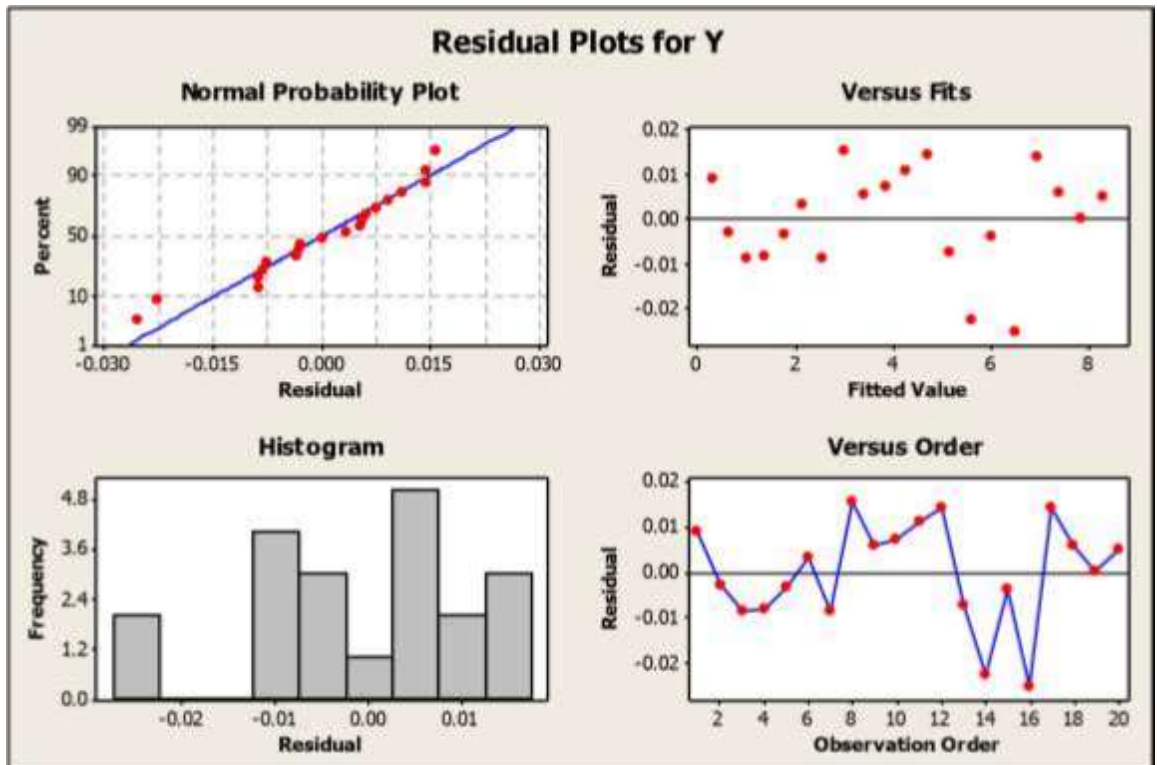
Figure (3): Residual plots for Y versus n, nlogn, n^2

*Interpretation of residual plots for Y:*
The normal probability plot suggests that the errors are almost normally distributed. The plot of residuals versus the fitted value of the response reveals that the distribution of the $\varepsilon_i$ has constant variance for all values of *n* within the range of experimentation. The plot of residuals versus observation order suggests that the errors are independently distributed, as there is no clear pattern in this plot.

### 3. CONCLUSION

Although the theoretical worst case complexity of RS sort is $Y_{worst}(n) = O(n\sqrt{n})$, the experimental results reveal that with $O_{emp}(nlgn)^{1.333}$ time complexity for typical inputs it can perform optimally. Interestingly, our algorithm, in average case could serve as a better choice for certain distribution data for which the popular quicksort algorithm is not an efficient choice. We leave the task of examining the behavior of RS sort for various continuous distribution inputs as a future work.

The general techniques for simulating the continuous and discrete as well as uniform and nonuniform random variables can be found in (Ross, 2001). For a comprehensive literature on sorting, see references (Knuth, 2001; Levitin 2009). For sorting with emphasis on the input distribution, (Mahmoud, 2000) may be consulted.

*System specification:* Below is the system specification.
Operating system: Windows 8 Pro (64-bit)
RAM: 4 GB

Hard Disk: 500 GB
Processor: Intel core i5, 2.5 GHz
Compiler: GNU GCC
Language: C++

Declaration:
"The authors declare that there is no conflict of interests regarding the publication of this article".